\documentclass[pre,preprintnumbers,amsmath,amssymb,floatfix,12pt,authoryear]{revtex4}

\usepackage{graphics}
\usepackage{amsmath}
\usepackage{graphicx,color}
\usepackage[latin1]{inputenc}
\usepackage[T1]{fontenc}
\usepackage{ae,aecompl}

\begin{document}

\vskip .27in

\begin{center}
{\Large \bf { Constraint optimisation and landscapes.}}

\vspace{.5cm}

Florent Krzakala$^1$ and  Jorge Kurchan$^2$,

\vspace{.5cm}
(1) PCT-ESPCI CNRS UMR Gulliver 7083 and (2) PMMH-ESPCI, CNRS UMR 7636,\\
 10 rue Vauquelin,
 75005 Paris, FRANCE

\end{center}

\vspace{1in}

\begin{narrowtext}

 We describe an effective landscape introduced in \cite{KK} for the 
 analysis of Constraint Satisfaction problems, such as Sphere
  Packing, K-SAT and Graph Coloring.
This geometric construction reexpresses
 these problems in the more familiar
terms of optimisation in rugged energy landscapes.
 In particular, it allows one to understand the
  puzzling fact that
 unsophisticated programs are successful  well 
beyond what was considered to be
the `hard' transition, and suggests an algorithm defining a
 new, higher, easy-hard  frontier.

\vspace{1cm}
PACS Numbers~: 75.10.Nr, 02.50.-r,64.70.Pf, 81.05.Rm

\end{narrowtext}

\pagebreak

Amongst  glassy systems, the particular class of `Constraint Optimisation'
has received constant attention \cite{giorgio,bookcomplexity}. 
These are problems in which we are given a set
of constraints that must be satisfied, and our task is to
optimize the conditions without violating them. The typical example is
packing: we are asked to put as many objects (spheres, say) in a given
volume, without violating the constraint that they should not overlap.
Another example that has been widely studied
by computer scientists is K-SAT, where we have $N$ Boolean variables,
and $\alpha N$ logic clauses: our  task is to add more and more clauses
while still finding some set of variables that satisfy them.
One last example is the $q$-coloring problem: we have a graph
with $N$ nodes and $\alpha N$ links and
our task is  to color each vertex with one of $q$ colours, with the condition
that linked vertices  have different colours. If we consider
a sequence of graphs 
as a set of nodes and a predefined list of links, 
then adding links one by one makes
the problem harder and harder. 

\begin{figure}[ht]
\begin{center}
  \includegraphics[width=10cm]{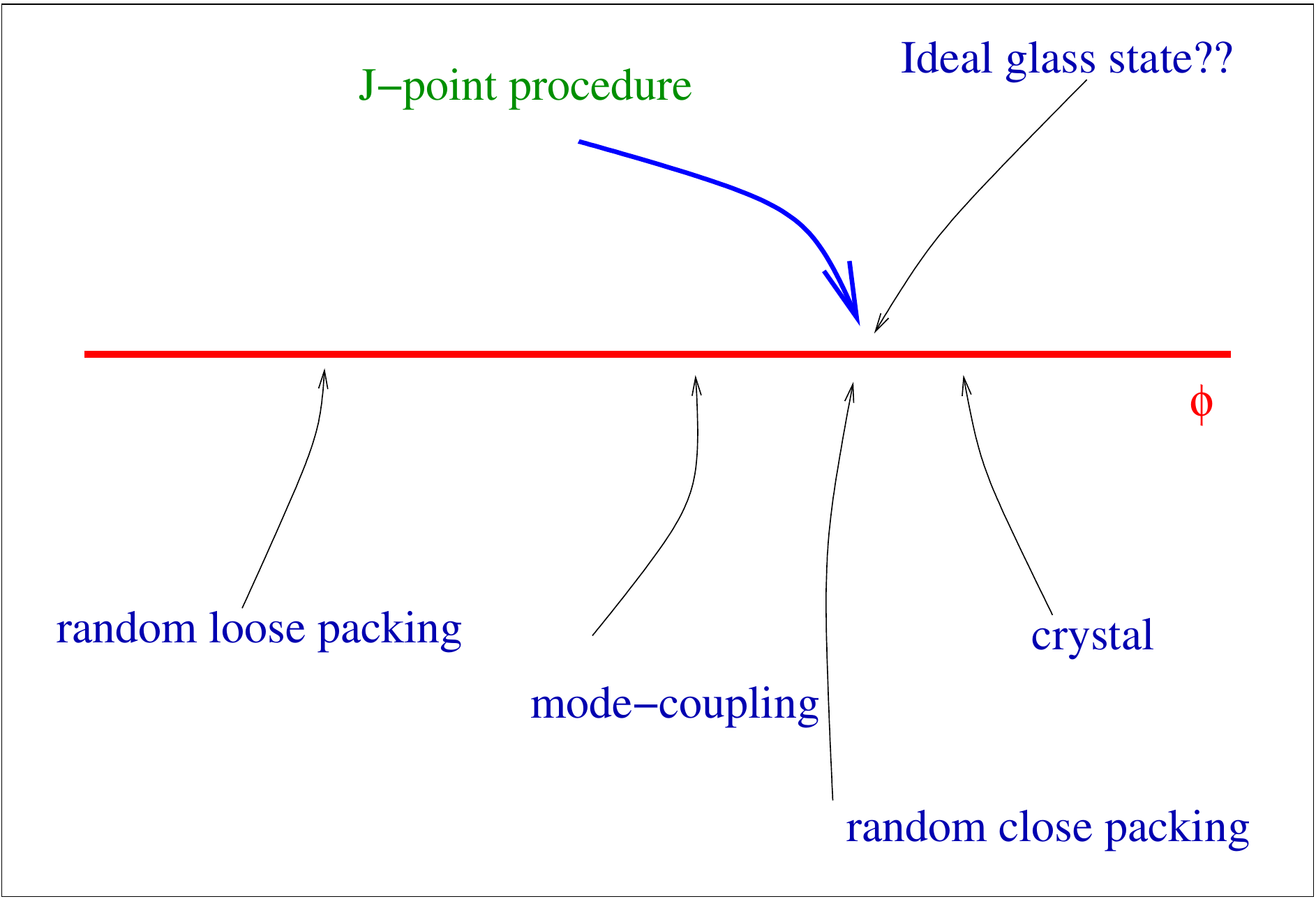}
\end{center}
\caption{The various transition densities 
for the sphere-packing problem.}
  \label{packing}
\end{figure}

What motivated our interest in these problems was what we percieved 
 as a confusing situation in the literature.
Consider  first sphere packing. In Fig. \ref{packing} we show different
volume fractions that are often quoted in the literature. In particular
`Random Close Packing' 
(as defined empirically),
the `optimal random packing' (zero-temperature glass state) and the so-called
`J-point', are sometimes used as synonyms and sometimes not.
The 'J-point' deserves some explanation. It can be  defined 
 as follows~\cite{OHern}: one starts from small spheres in random
 positions, and `inflates' them gradually (in the computer, of course), 
only displacing them the least neccessary to avoid overlaps~\cite{LS}.
At some point the system blocks and the procedure stops: this is the J-point. 
It was studied extensively by  the Chicago 
group~\cite{OHern,TheseM}, who 
proposed that it  be 
identified with Random Close Packing. On the other hand,  Random Close Packing
is often associated with the  zero-temperature ideal glass state. The two
 identifications seem hardly compatible, as they would imply that the 
fast algorithm described above allows to find rapidly the ideal glass state,
contrary to all our prejudices.

Let us now turn to the SAT and Colouring problems. Carrying over the 
knowledge from mean-field glasses, it  was concluded that the set
of solutions evolves, as the difficulty is increased, in the following
 manner:
for low $\alpha$ the set of solutions is connected. As $\alpha$ is increased 
there is a well defined `dynamic' or `clustering' point $\alpha_d$ at
which the set of satisfied solutions breaks into many comparable 
disconnected pieces \cite{uff}. At a larger value $\alpha_K$ the volume becomes
dominated by a few regions, and finally, at some $\alpha_c$, there are no more 
solutions~\cite{all}.

\begin{figure}[ht]
\begin{center}
  \includegraphics[width=11cm]{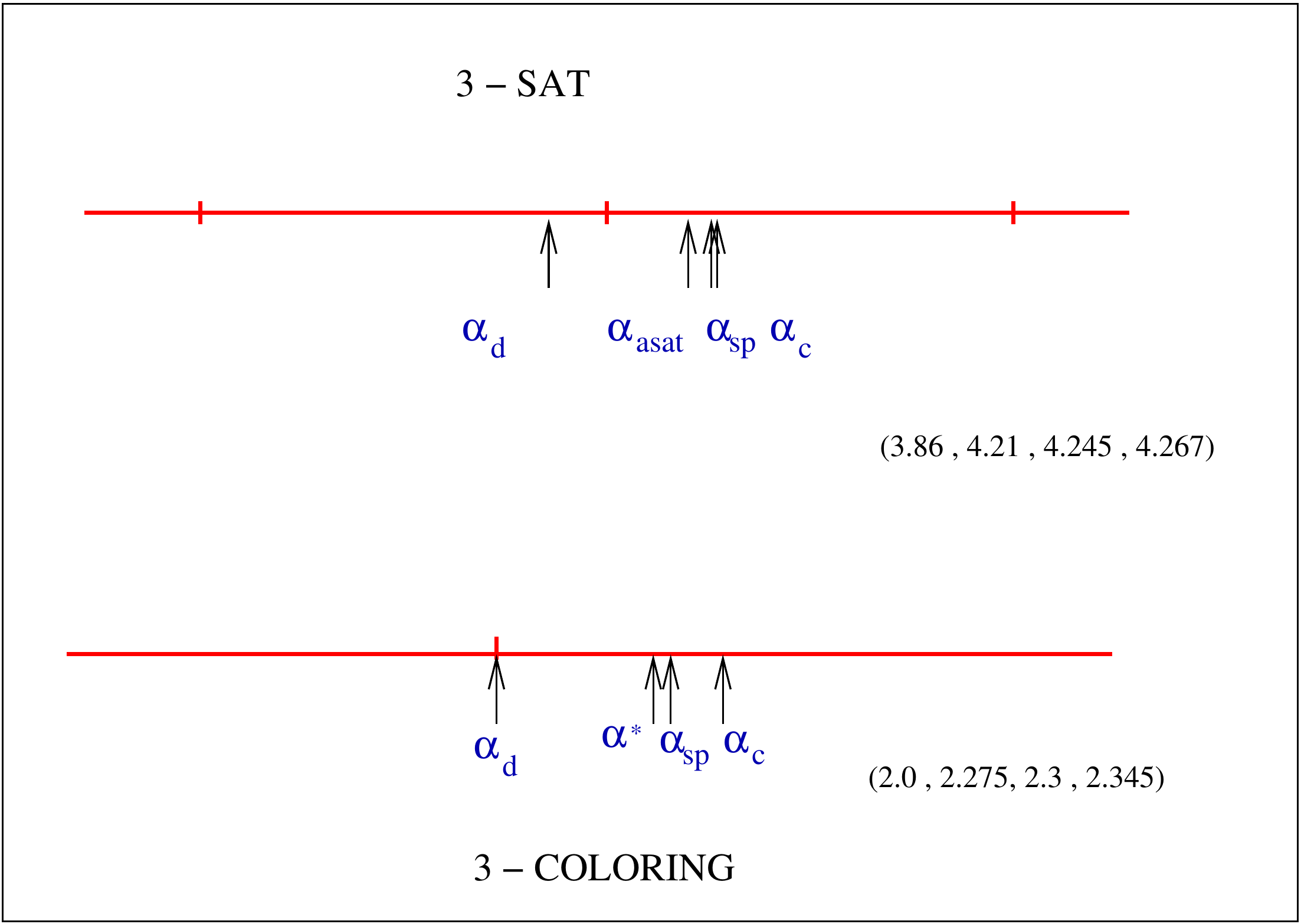}
\end{center}
\caption{Why is it so easy to go beyond $\alpha_d$, 
the putative {\em `hard'} limit?
Values of the parameter: i) $\alpha_d$ the `clustering' transition,
ii) $\alpha_{ASAT}$ for ASAT, $\alpha^*$ for our algorithm,
 iii) $\alpha_{SP}$ the performance of a Survey Propagation implementation, and
iv) $\alpha_c$ the optimum~\cite{alpha}.
}
  \label{alphas}
\end{figure}

Beyond the clustering transition $\alpha_d$  the problem  
was thought to become {\em hard}. 
{\em And yet, as it turned out, even very simple programs\footnote{
Here by `simple' or `unsophisticated' we mean `using no specific knowledge of
the glass state'.} 
 manage to find solutions well beyond this hard transition!} 
The situation is showed in  Fig. \ref{alphas}.
 This is another puzzle we set out to clarify.

\begin{figure}[ht]
\begin{center}
  \includegraphics[width=10cm]{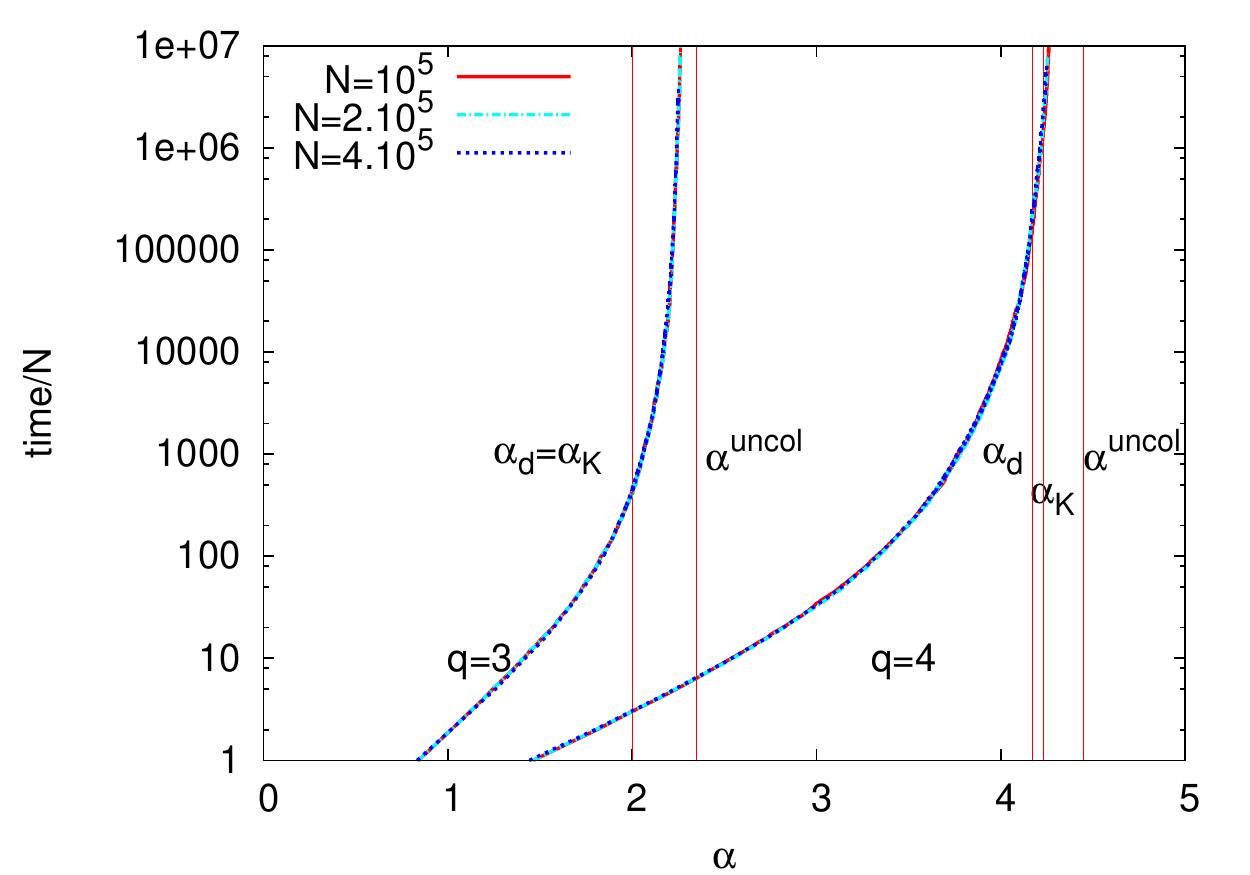}
\end{center}
\caption{Integrated number of colour flips needed to avoid miscolourings,
per unit size, for the three and four colouring problem. The
clustering transition and glass transitions are no obstacle.  }
  \label{divergent}
\end{figure}

A first observation one can make 
 is that the `J-point' procedure can be generalised 
to all of these problems: one just has in all cases to
 increase the difficulty gradually,
and keep the system satisfied by minimal changes each time. For example,
for the Colouring problem, one adds one link at a time, and corrects any
miscoloring generated by such an addition. The number of colour 
flips needed each
time to correct the miscoloring grows and it diverges with a well-defined,
reproducible
power law (see Fig. \ref{divergent}) at a value $\alpha^*$, by definition
the limit reached by the program.

\begin{figure}[ht]
\begin{center}
  \includegraphics[width=8cm]{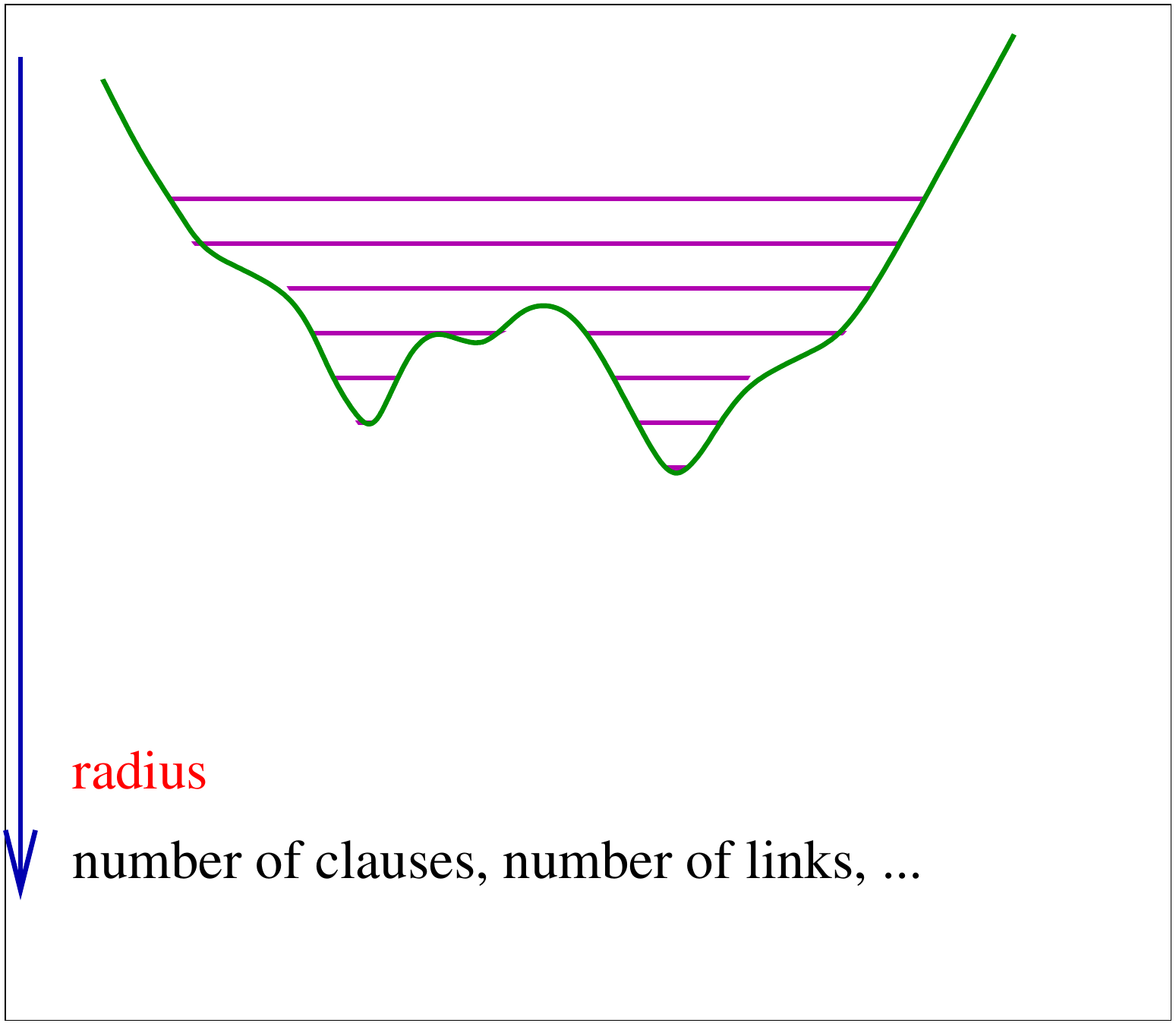}
\end{center}
\caption{Pseudo-energy (conjugated to pressure) landscape for constraint
optimisation problems. The sets of satisfied configurations at increasing
levels of difficulty are included in the previous: this allows for 
the definition of a 
well-defined envelope.}
  \label{pseudo}
\end{figure}

Second, and most important, we introduce a (pseudo) energy landscape
as follows (see Fig. \ref{pseudo}). As the difficulty in the problem
is increased -- by increasing the radius, or adding clauses, or adding links --
the set of satisfied configurations becomes a subset of the previous one.
This allows to construct a single-valued envelope function (Fig. \ref{pseudo}):
the pseudo-energy.
{\em It is easy to see that the J-point procedure is just a zero-temperature
descent on this landscape}. 

We can now carry over everything we know from energy landscapes.
For the J-point in the context of sphere packings, we  conclude
that:
\begin{itemize}
\item The J-point, being the result of a gradient from a random configuration,
cannot be the optimal amourphous solution. It is just the analogue of
the {\em infinite temperature inherent structures.}
\item It is in general more compact than the clustering (Mode Coupling)
point,
 since it gains from `falling to the bottom of one cluster'.
\item
It may be more or less compact  than the Kauzmann ($\alpha_K$) 
 point itself, depending on the dimension, polydispersity, shape, etc.
\end{itemize}

For problems such as SAT and Coloring, we have  now a {\em 
recursive incremental } algorithm, in which one increases the difficulty at
small steps, at the same time correcting the configuration minimally in order 
to stay satisfied~\cite{foot1}.
 This algorithm finds solutions in polynomial time up to a $\alpha^* \geq
\alpha_d$. 
Once $\alpha_d$ is reached, the current solution remains `trapped'
 within one
cluster. On increasing further $\alpha$
(for example, in the Coloring problem, by adding  further links),
 the cluster of solution contracts
until it finally dissappears at $\alpha^*$. 
As one can see in Fig. \ref{alphas}, one can go quite a long way beyond
$\alpha_d$. How much so  depends  on
how fast the cluster dissapears: gradually for small $q$ and $K$ 
(in Coloring and SAT, respectively), and  essentially immediately for large
$q,K$ and for problems that have variables whose value is
 frozen within a cluster.

Our conjecture is that unsophisticated programs will not do better than
$\alpha^*$, or rather, than its `slow annealing' version as 
above \cite{foot1}.  
Comparison with message-passing 
algorithms such as Belief and Survey Propagation is complicated by the fact 
that neither our version of the Recursive Incremental program, nor
the published implementations of Survey Propagation have been pushed to their
 optimum~\cite{foot2}. With this caveat, the Survey Propagation algorithm
 seems to do better
in the Coloring problem. On the other hand,
Braunstein and Zecchina have recently shown that a message-passing algorithm
does well in the Binary Perceptron model~\cite{BZ} -- a problem with 
single-configuration states -- and this suggests
that these algorithms  go beyond $\alpha^*$ in that case. 
At any rate, it would be very interesting to explore along these lines 
the $K=3$
SAT problem, a much better studied case.

Perhaps the greatest promise of this approach comes from the fact that,
as we have indicated in Ref. \cite{KK},  $\alpha^*$  defined by the
Recursive  
Incremental algorithm lends itself,  due to its simple geometric
definition,
 to an analytic computation.


\begin{thebibliography}{99}
\bibitem{foot1}One can also envisage a further improvement, in analogy with the
 Lubachevsky-Stillinger \cite{LS} algorithm: one can follow
each  increase in $\alpha$ with a number
of  steps of diffusion between satisfied
solutions.
\end{thebibliography}
\end{document}